\documentclass[12pt,preprint]{aastex}
\usepackage[usenames,dvips]{color}
\def\deg{^{\mathrm o}}

\newcommand{\swift}{{\it Swift}}
\newcommand{\xmm}{{\it XMM-Newton}}

\slugcomment{submitted to AJ \today} \shorttitle{Absorption in Mrk
1393} \shortauthors{Wang et al.}
\begin{document}
\title{X-ray Absorption and Optical Extinction in
the Partially Obscured Seyfert Nucleus in Mrk 1393}
\author{T. G. Wang\altaffilmark{1,5}, H.Y. Zhou\altaffilmark{1,2},
D. Grupe\altaffilmark{3},
W. Yuan\altaffilmark{4},
X.B. Dong\altaffilmark{1,5}, H.L. Lu\altaffilmark{1,5}}
\altaffiltext{1}{Center for Astrophysics, University of Science and Technology of
       China, Hefei, Anhui, 230026, P.R.China}
\altaffiltext{2}{Max-Planck-Institut f\"ur extraterrestrische Physik, Postfach 1312, 85741 Garching, Germany}
\altaffiltext{3}{Department of Astronomy and Astrophysics, Pennsylvania State University, 525 Davey Lab, University Park, PA 16802}
\altaffiltext{4}{National Astronomical Observatories/Yunnan Observatory, Chinese Academy of Sciences, Kunming, Yunnan 650011, China}
\altaffiltext{5}{CAS Key Laboratory for Research in galaxies and Cosmology, University of Science and Technology of China, Hefei, Anhui, 230026, China}

\begin{abstract}

We present a detailed study of the X-ray and optical spectra of the
luminous Seyfert galaxy Mrk 1393, which revealed variable partial
obscuration of the active nucleus. The X-ray spectra obtained by
\xmm~and \swift~show moderate absorption with a column density
around 3$\times 10^{21}$~cm$^{-2}$, consistent with a dust-reddening
interpretation of the steep Balmer decrement seen in recent
optical spectra. The X-ray flux in the 0.5 to 2 keV band 
during the \xmm\ observation in 2005 and \swift~observation in 2006 was 
a factor 6 brighter than that of the ROSAT All Sky Survey in 1991. In 
the past 4 years, the broad H$\alpha$ line brightened by a factor of 4 
accompanied by a decrease in the Balmer decrement. A comparison with 
literature spectra reveals 
variations in the dust extinction on time scales of several years, suggesting
that the obscuring material is very close to the active nucleus.
These observations indicate that a dust-to-gas ratio as high as the
Galactic value can be present in moderately thick gas in the
vicinity of the central engine within a few parsecs. We suggest that
the obscuring material may be debris disrupted from the dusty torus.

\end{abstract}

\section{Introduction}

The unification scheme for type I and type II nuclei of Seyfert
galaxies has been very successful in explaining various
observational facts: detection of polarized broad emission lines in
a significant fraction of Seyfert 2 galaxies, large X-ray absorption
column densities and/or reflection dominated X-ray spectra in
Seyfert 2 galaxies \citep*[e.g.][]{vv00}, presence of ionization
cone and a torus-like near-infrared image \citep{mason06,jaffe04}.
However, we still do not understand the properties of the torus.
Clumpy distribution is preferred by recent modeling of high
resolution infrared image of nearby Seyfert galaxies and the
infrared spectral energy distribution \citep*[e.g.][]{hon08}, but
yet little is known about the properties of dust clouds, such as
their dust-to-gas ratio. Furthermore, it is highly controversy
whether the subtending angle of dust torus varies with nuclear
luminosity \citep[][]{steffen03,bar05, hao05}, \citep*[c.f.,]
[]{mar06,eck06, wang07}.

Partially obscured active galactic nuclei (AGNs) are ideal targets
for exploring some of these questions. First, since the transmitted
light can be directly measured, the estimation of the
reddening/absorption column density is much more reliable than in
type II quasars \citep[e.g.][]{mai01a}. This allows a reliable
measurement of the dust-to-gas ratio of the obscuring material. The
transmitted light also allows us to measure properly the nuclear
properties \citep*[e.g., intrinsic nuclear variability,][]{dong05},
and puts better constraints on the other parameters (such as the
size of the broad emission line region). Second, optical and X-ray
observations also show that at least some of these partially
obscured AGNs display transitions from type 1.8/1.9 to 1.2/1.5 or
vice-versa \citep*[e.g.][]{goo89}. Two different scenarios are
proposed to explain these transitions: a continuum luminosity
dependent Balmer decrement variation \citep[e.g.,][]{wam90,tran92,shap04}, 
and the dust absorbing
material moving in/out of line sight. Clearly, simultaneous X-ray
and optical observation can give important constraints of these
models. If the latter scenario is confirmed, these transitions carry
important information of the distribution and kinematics of
absorbing material.

We initiated a program to observe partially obscured AGNs and
quasars in X-rays with the \xmm\ Observatory \citep{jansen01}. The
targets were selected from the Sloan Digital Sky Survey (SDSS, York
et al. 2000) by their large Balmer decrements \citep*[][]{dong05}.
In this paper, we report of the detections of X-ray absorption and large
variations of the Balmer decrement in one of these objects, Mrk 1393
($z=0.0544$). The X-ray observations were performed by  \xmm\ and
\swift. Mrk 1393 appears to be a partially obscured Seyfert galaxy
due to its Balmer decrement as large as H$\alpha$/H$\beta=7.7$
measured from  its SDSS spectrum. Mkn 1393 is a Seyfert 1
galaxy with prominent broad H$\beta$ \citep{kinman83}, and is also
included in the broad line AGN sample of Hao et al. (2005). The
reddening corrected H$\alpha$ luminosity of 10$^{42}$ erg~s$^{-1}$ makes this
AGN a borderline object between Seyferts and QSOs.
It is also infrared luminous with $M_K=-25.03$ and was detected by
ROSAT in X-ray (Anderson et al 2003). With a radio flux of 23~mJy 
detected in the FIRST survey (Becker, White \& Helfand 1995), it is 
formally radio loud with an $R_r=\log (f_{1.4GHz}/f_r)=1.42$ (Ivezic 
et al. 2002). The paper 
is organized as
follows: the X-ray data obtained by \xmm\ and Swift are analyzed in
\S2. The optical spectroscopic observations and analysis are
presented in \S3. The properties of the obscuring material is
discussed in \S 4. Throughout this paper, luminosities are
calculated assuming a $\Lambda$CDM cosmology with $\Omega_{\rm
M}$=0.27, $\Omega_{\Lambda}$=0.73 and a Hubble constant of $H_0$=72
km s$^{-1}$ Mpc$^{-1}$, corresponding to a luminosity distance of
D=236.2 Mpc to the galaxy.

\section{Results from X-ray data}

\subsection{\xmm\ and \swift\ observations and data reduction}


Mrk 1393 was observed by \xmm\  on 2005 July 20 with the EPIC PN
\citep{strueder01} and MOS1+2 \citep{turner01} for 4423s, 6565s, and
6570s, respectively. The PN and the MOS detectors were operated in
full-frame mode with the thin filter. Due to some strong background
flares at the beginning of the observation, part of the data had to
be excluded from further analysis. 
After excluding the time intervals with a hard
X-ray background ($E>10$ keV) of more than 10 counts s $^{-1}$, 
the net observing times are 4099s, 5324s, and 5335s for the PN,
MOS-1, and MOS-2, respectively. The \xmm\ data were reduced with
{\em XMMSAS} 7.1.0. Only single and double events were used for the
PN data analysis, and single to quadruple events for the MOS data
analysis. Data were also selected with the quality flag set to {\tt
FLAG=0}. The source counts are extracted from a circular region with
a radius of 90$\farcs$ around the source position. Background
counts were accumulated in a circular region of the same radius close by.
The net source count rates in the 
0.2-10 keV band are
2.087 cts~s$^{-1}$ for the {\em PN} detector and 1.577 cts~s$^{-1}$
for {\em MOS1+2} detectors, indicating that the data are not
significantly affected by pileup. The light curves were binned in
100s bins. The X-ray spectra were rebinned using {\it grppha} version
3.0.0 with at least 20 counts per bin. The redistribution and
auxiliary response matrices were created by the {\em XMMSAS}  tasks
{\em rmfgen} and {\em arfgen}, respectively.

During the \xmm\ observation, Mrk 1393 was observed with the
on-board Optical Monitor \citep[OM; ][]{mason01} with the UVM2
filter. The OM data were reduced with the {\em XMMSAS} task {\it
omichain}. The measured UVM2 magnitude from the source list is
M2=16.17$\pm$0.08 mag. Correcting for the Galactic reddening of
$E_{\rm B-V}$=0.059 mag \citep[][]{schlegel98} results in M2=15.60
mag.

Mrk 1393 was  observed by \swift~\citep[][]{gehrels04} on
2005 September 10  for a total exposure time of 9.6 ks. The
\swift-XRT \citep{burrows05} observation was performed in photon
counting mode \citep{hill04}. The event file of the observation was
created by using the \swift\ analysis tool {\it xrtpipeline} version
0.11.4. Source photons were extracted in a circle with a radius of 47\farcs~ 
centered on the source, 
and the background was selected from a source-free region 
with a radius of 188\farcs. Only
single to quadruple events in the energy range of 0.3-10.0 keV were
selected for further analysis. Source and background spectra were
extracted from the event file by using {\em XSELECT} version 2.3.
Spectra were rebinned within {\it grppha} 3.0.0 to have at least 20
photons per bin. The auxiliary response files were created by the
\swift\ tool {\it xrtmkarf}. We used the response matrix version 010
with a grade selection 0-12. The data are not significantly affected
by pileup due to the low count rate ($\sim$0.2 cts~s$^{-1}$). The XRT 
light curve was created
by a program in MIDAS\footnote{MIDAS = ESO's Munich Image Data Analysis System}
having at least 200 photons per bin (see e.g.
Nousek et al. 2006 for details).

In addition to the X-ray data, we also obtained photometry with the
UV/Optical Telescope \citep[UVOT; ][]{poole08} in the UVM2 filter.
Source photons were extracted from a circular region with
r=5$^{''}$, and the background in an annulus around the source with
an inner radius of 7$^{''}$ and an outer radius of
20$^{''}$\footnote{At the SDSS u-band, the radius containing 50\% of
Petrosian flux is about $2^{''}.0$.}. The UVOT tool {\it uvotsource}
was used to determine the magnitudes and fluxes. We searched for UV
variability in the individual segments of the observation and did
not detect any. The average observed magnitude based on the co-added
images is m$_{\rm UVM2} = 15.88\pm0.01$ mag (corresponding to a flux
of $f = 2.11 \times 10^{-15}$ ergs s$^{-1}$ cm$^2$ \AA$^{-1}$
corrected for Galactic reddening \citep[][]{schlegel98}. This value
is consistent with the UVM2 OM magnitude of $M2_{adj}=15.90$
mag\footnote{For a comparison with the \swift-UVOT, the UVM2 OM
magnitudes have to be adjusted by +0.30 mag \citep{grupe07,grupe08}.},
indicating no significant change in the UV flux between the \xmm\
and \swift\ observations.

\subsection{X-ray Spectral Analysis}

The X-ray spectra are fitted by using XSPEC
\citep[][v.12.3.1x]{arnaud96}. Errors of derived parameters are
quoted at $\Delta\chi^2$=2.7. The  Galactic
HI column density in the direction of the source is
4.8$\times$10$^{20}$ cm$^{-2}$  \citep{kal05} and is
always taken into account in  all the spectral fits as listed in Table\,1.

The \xmm~ spectra in the 1-10 keV energy band can be fit by a
power law model
with additional intrinsic neutral absorption. 
The intrinsic absorption column density at the resdshift of the AGN is
 (2.4$\pm$0.6) $\times$ 10$^{21}$~cm$^{-2}$. The
X-ray spectrum in the range 1-10 keV is somewhat flat
($\Gamma=1.50\pm 0.06$), even for typical Seyfert 1 galaxies
\citep[e.g. ][]{leighly99, grupe01}. There is no apparent
iron K$\alpha$ line at the corresponding energy. The upper limit to a narrow
($\sigma=0.1$ keV) iron K$\alpha$ line is 30 eV.

A flat spectrum may result either from partial covering
absorption\footnote{Partial covering may be caused by a collection
of absorbing blobs which are individually smaller than the X-ray
emitting region, such as broad line emission clouds.} or from strong
reflection (reprocessing) \citep[e.g.][]{fabian89,gallo06,grupe08}. 
Here we examine both possibilities. A
reflection model ({\em pexrav} in {\em XSPEC}) does not improve the
fit, yielding a best fitted normalization $R$ of the reflection
component of zero and an upper limit of $R<1.52$ (fixing the
inclination to 45$\deg$ and the metal abundance to the solar value).
This is consistent with the upper limit derived from the equivalent
width of the neutral iron line. An excess absorption and a flat
photon index are still required, with the same values as the simple
power law fit ($\Gamma=1.44_{-0.05}^{+0.14}$ and
$N_H=2.2_{-0.5}^{+0.9}$\,10$^{21}$\,cm$^{-2}$). Neither does a
partial covering model improve the fit. The covering fraction
(0.4--1.0), though consistent with a full coverage, is poorly
constrained, and so is the absorption column density
($N_H=3.8^{+8.4}_{-2.1}\times10^{21}$\,cm$^{-2}$). The photon index
remains flat ($\Gamma=1.48^{+0.13}_{-0.07}$). We thus conclude that
the excess absorption and a flat continuum index inferred from the
1--10 keV spectra are caused by neither partial covering nor
reflection.

\begin{figure}
\epsscale{0.8}
\plottwo{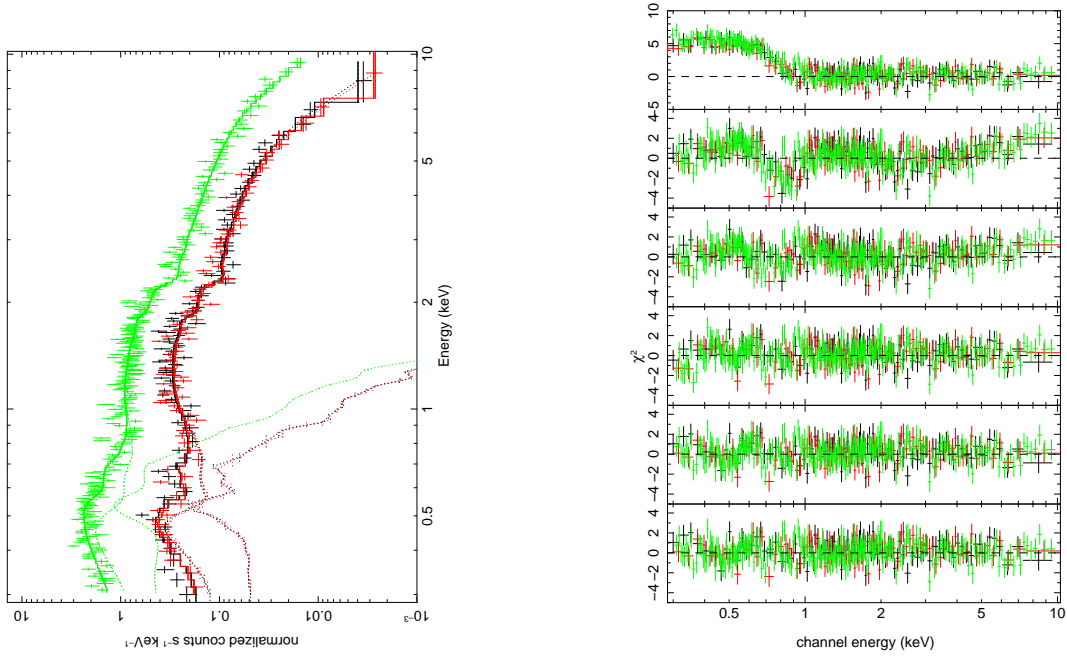}{fig1b.ps}
\caption{Left panel: The best fit to
the \xmm\ spectrum in the 0.3-10 keV band with a model consisting of
a broken power-law (short-dashed) absorbed by cold gas and thermal plasma
emission (dotted) with the PN data in green and MOS data in red and black. 
The individual components of the best-fit
model are also shown. Right panels: residuals for different models
(from top to down): a) powerlaw model extrapolated from 2-10 keV;
(b) partial covering absorption model; (c) single ionized absorption 
model (absori); (d) absorbed powerlaw plus {\em meka} model; (e) 
power-law+black body emission absorbed by cold gas plus{\em meka} model; 
(f) broken power-law absorbed by cold gas plus {\em meka} model} \label{xspec}
\end{figure}

We now consider the whole 0.1--10 keV energy band. Extrapolating the
1--10 keV best-fit power law model down to lower energies
under-predicts the soft X-ray flux by a factor of more than three
below 0.6 keV, suggesting the presence of a possible strong soft
X-ray excess. It is known that in some cases ionized absorption
models can mimic soft excess (e.g., Komossa \& Meerschweinchen 2000,
Done et al. 2007). We first consider an ionized absorption model to
examine this possibility. We fit the simple model {\em absori}
available in {\em XSPEC}. This produces a marginally acceptable fit 
with a $\chi^2$=703 for 652 d.o.f. (corresponding to a probability level
of $P_r\simeq 8\%$). The absorber has moderate ionization (with an ionization 
parameter $\xi=L_x/4\pi R^2 n\simeq18_{-3}^{+4}$ erg~cm~) and is 
considerably thicker ($N_H=1.0\pm0.1\;10^{22}$~cm$^{-2}$) 
than in the cold absorption model. The fit also yields significantly 
steep X-ray spectrum ($\Gamma=1.79\pm 0.03$). However, systematic 
deviations are still seen in the residuals both in 0.5 to 1.5 keV range 
and above 5 keV (see \ref{xspec}). The systematic deviations may be 
attributed partly because the  
{\em absori} model does not fully represent the spectrum, and 
partly to the multi-phase of the ionized absorber. In order to examine 
the latter possibility, we add a second {\em absori} to the model. 
This results in a $\Delta\chi^2=46$ for two more free parameters, 
and all systematical residuals below 2 keV disappear now. The two 
components have ionization parameters $\xi_{low}=4.5_{-1.2}^{+2.1}$ 
and $\xi_{high}=259_{-114}^{+281}$, and column densities 
5.4$_{-0.9}^{+1.0}$~10$^{21}$~cm$^{-2}$ and 
2.6$_{-1.0}^{+2.0}$~10$^{22}$  
for 'low' and 'high' ionization phases, respectively.

Next, we consider the soft X-ray excess to be a real emission feature 
rather than induced by ionized absorption, and add
a soft X-ray component in the spectral fit. Adding soft excess
emission often causes degeneracy between the strength of soft excess
and the amount of absorption. Since X-ray absorption is our main
concern in this work, we focus the following analysis on detecting
and measuring the amount of any possible X-ray absorption, by
testing various spectral models of the soft excess, namely,
black-body, ionized plasma, and power-law.

One of the common spectral shapes reproducing the soft X-ray excess
emission in AGN is a black-body. We first test a power-law plus a
black-body model, both absorbed by the same amount of
(cold) absorption. However, the fit is hardly acceptable
($\chi^2/d.o.f=720/657$), with a probability level of only
$P_r\simeq 3\%$. The fitted absorption column density and photon index are
$1.9_{-0.3}^{+0.3}\times10^{21}$~cm$^{-2}$ and
$\Gamma=1.43_{-0.05}^{+0.04}$, respectively. We also use a broken
power-law to account for the spectral steepening at the low
energies, and find it unacceptable either, with $P_r\sim 5\%$.

Soft X-ray  lines are commonly seen
in Seyfert 2 and intermediate Seyfert galaxies, which is interpreted
as arising from extended photoionized gas, such as the narrow line
region (NLR). This component is less prominent in Seyfert 1 galaxies
because of strong nuclear X-ray emission. In the case of Mrk 1393,
however, this component may become prominent given the likely strong
attenuation of the  nuclear continuum below 1 keV (see above). Here
we use a thermal plasma model ({\em meka} in XSPEC), which is a
reasonable approximation of the spectrum of photoionized gas given
the spectral resolution and signal-to-noise ratio of the current
X-ray observations. We leave abundances and temperature as free
parameters. Acceptable fits can be achieved only when the plasma
emission is free from the excess absorption required for the
power-law continuum. This is consistent with the plausible nature of
as being extended emission. This model yields an acceptable fit with
$\chi^2/d.o.f=666/650$ ($P_r=32\%$), and a similar absorption column density 
$N_H=3.0_{-0.5}^{+0.5}\times 10^{21}$~cm$^{-2}$, and a flat
continuum $\Gamma=1.53_{-0.05}^{+0.05}$, $kT=0.166\pm 0.004$ keV,
and a low abundance of $Z/Z_\odot=0.023_{-0.005}^{+0.006}$ (see
Table\,1). The low abundance suggests that this component is 
dominated by continuum rather than emission lines. Thus, it raises 
a big concern for interpreting UT as from an photoionized gas  
\footnote{Note that a similarly low abundance was obtained
in a large number of other sources; for instance, one example is the
Seyfert 2 galaxy NGC 5252 with thermal emission model
\citep[][]{cap96}. This commonly seen effect is generally traced
back to multiple components or non-equilibrium effects in the
emitting plasma.}. 
The total luminosity of this component is $\sim
4\times 10^{42}$~erg~s$^{-1}$ in 0.5--2.0 keV, higher than those of
most Seyferts but comparable to the luminous extended X-ray emission
in NGC 6240 (Komossa et al. 2003). This large soft X-ray luminosity
we observe is unexpected because the [OIII] to soft X-ray flux ratio
in Mrk 1393 is 20 times lower than the typical values found in
Seyfert 2 galaxies \citep{cap96}. So it is likely that this soft
X-ray emission component in Mrk 1393 is diffuse, possibly
contributed by photoionized gas (e.g.\ the inner NLR) and/or thermal
gas, or scattered light of the continuum.

Though an absorbed power-law plus an unabsorbed {\em meka} component
gives an acceptable fit, there are systematic residuals in the soft
energy band, however. In fact, we find that adding a second soft
X-ray component with the same absorption column density
significantly improves the fit at low energies. Using a
black body model as the (nuclear) soft X-ray component reduces the
$\chi^2$ by $\Delta\chi^2=23$ for 1 more free parameter (643/649). 
The parameters of the best-fit are listed in Table\,1, where the
abundance of the {\em meka} model is fixed at the solar value. 
The solar abundance also ensures that this component can be interpreted
as emission from photoionized gas. There
are little changes in the column density ($2.4_{-0.3}^{+0.4}\times
10^{21}$~cm$^{-2}$) and the photon index ($1.48_{-0.04}^{+0.05}$).
For the plasma emission ({\em meka}) the
 temperature remains unchanged (0.167$\pm0.005$ keV),
while its luminosity is reduced by $\sim40\%$ compared to the
previous value (2.6$\times 10^{42}$~erg~s$^{-1}$). Using a broken
power-law model instead of the power-law plus blackbody model 
yields a similarly
good fit and similar model parameters (see  Table\,1), with a
luminosity of 2.2$\times\ 10^{42}$~erg~s$^{-1}$ in 0.5--2.0 keV band for
the {\em meka} component.

Based on the above results, extensive fits with various plausible spectral models,
we find that the amount of 'cold' X-ray absorption is surprisingly robust, 
independent of the choice of spectral models in the soft X-ray band.
We conclude that, for Mrk 1393, the excess 'cold' X-ray absorption is 
inevitably required, and its measured column density is relatively 
insensitive to the choice of the spectral shape of the soft X-ray excess 
components. Both the soft excess and excess absorption can also be 
accounted by two-phased ionized absorption medium. 

The {\rm SWIFT} XRT spectrum has a lower signal-to-noise ratio compared
to the \xmm\ spectrum. We simply apply those best-fit models from
the \xmm\ data to the \swift\ spectrum. We find that all these
models fit the \swift\ spectrum quite well, yielding  almost the same
model parameters as those derived from the \xmm\ data. Therefore,
the two spectra are well consistent with each other. In Table\,1, we
list results of only one of the best model fits for demonstration.

To summarize, the \xmm\ spectrum requires intrinsic absorption
of `cold' gas with a column density $N_H\simeq 2.2-3.3 \times 
10^{21}$~cm$^{-2}$. The power-law spectrum above 1 keV has a photon 
index of $\Gamma\simeq 1.4 -1.5$, marginally flatter than typical 
Seyfert 1 galaxies. An additional emission component, probably a combination
of scattered light and emission lines from collisionally
or photoionized gas, is inferred to be present
with a luminosity of a few times $10^{42}$~erg~s$^{-1}$ in 0.5--2.0 keV
and an equivalent temperature of 0.16 keV. These parameters do not vary 
much for the models considered here. Alternatively, a two phased ionized 
absorber may account for both the excessive absorption and the soft X-ray 
excess. The amount of 'low' ionization phase medium is a factor of two 
larger than that in the cold absorption model.

\subsection{X-ray variability}

\begin{figure}
\epsscale{0.8} \plotone{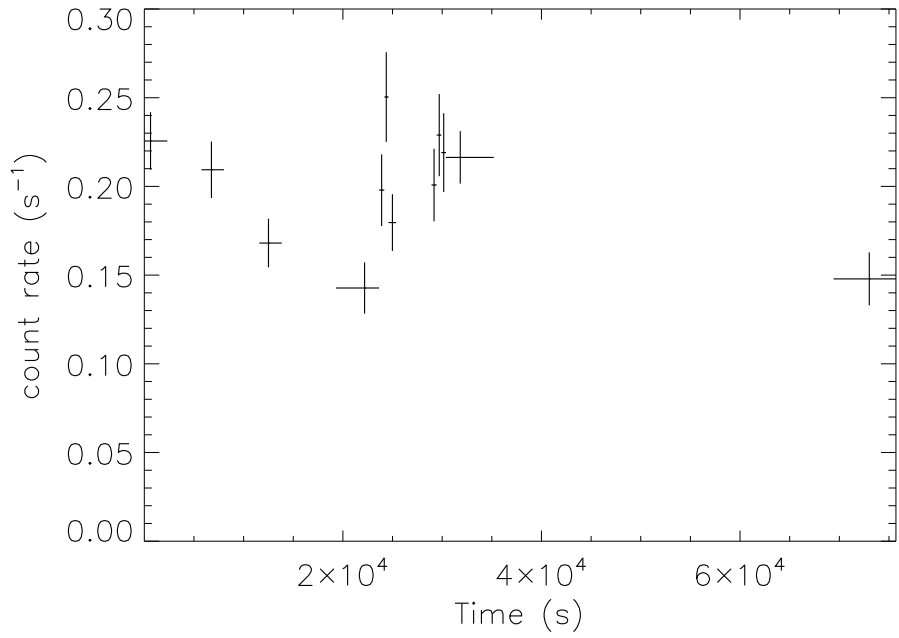} 
\caption{The background-subtracted X-ray light curve during Swift XRT 
observation. The curve is rebined to have 100 counts per bin. Short 
time scale variability is apparent.} 
\label{swift_curve}
\end{figure}

The \xmm\ observation lasted less than 7 ks, during which no
significant X-ray variability is found. To search for variability
in the \swift\ observation covering 19 hr in total, the count 
rates are corrected according to the exposure maps to compensate losses
induced when the source was positioned on bad CCD columns. The
light curve (Fig \ref{swift_curve}) shows some variability on the 
order of 20\% within an hour. The curve is rebined that each bin 
contains 100 counts. A $\chi^2$-test gives $\chi^2=41.8$ for 11 degree 
of freedoms, which has a null probability of less 2$\times10^{-5}$. 
Furthermore, we also compare the average \swift\ fluxes in the 0.3-1.0 
and 2.0-10.0 keV bands with the XMM-Newton measurements. The mean 
fluxes do not show any significant variability on month timescales.

Mrk 1393 was detected during the ROSAT All Sky Survey \citep[RASS;
][]{voges99} with a count rate of 0.033$\pm$0.012 cts~s$^{-1}$ in
the 0.1-2.4 keV band and a hardness ratio of HR1=(H-S)/(H+S)=0.73$\pm$0.43
with the H and S being the counts in the hard (0.5-2.0 keV) and the soft 
(0.1-0.4 keV) bands, respectively \citep{voges00}. The observation was 
made between August 6 and August 8, 1991.  For comparison, we estimate 
the ROSAT count rate for the best-fit \xmm\ model using the ROSAT PSPCC 
response. The broken power-law plus {\em meka} model predicts an RASS 
count-rate of 0.189 cts~s$^{-1}$ in the 0.5-2.0 keV band. The power-law 
plus black body soft excess and {\em meka} model gives a very similar 
count-rate. With the observed hardness ratio and count-rate in the whole band, 
we estimate the observed RASS count rate in the 0.5-2.0 keV band to be 
0.028 cts~s$^{-1}$. Therefore, the soft X-ray flux increased by 
a factor of ~6 since the ROSAT observation if we ignore the inter-mission 
calibration uncertainty, which is estimated to be within 10\% 
\citep{snowden02}.  In passing, we note that the variation of intrinsic 
X-ray luminosity might be larger than this apparent value. Thermal 
plasma emission with similar temperatures has been observed in Seyfert 2 
galaxies, which is emitted on relatively large scales, and is likely not
variable on time scale less than ten years \citep*[e.g.][]{bgc06}.
The {\em meka} component alone predicts a ROSAT count rate in the band 
(0.5-2.0) keV of 0.027 cts~s$^{-1}$ with 15\% uncertainty due to its 
normalization, very close to the observed RASS count-rate. If the {\em 
meka} component comes from the emission from the extended photoionized 
gas, it might be not variable. Thus the actual variability amplitude 
can be even larger.

\section{Optical Observations and Emission Line Measurements }

\subsection{Spectra Obtained at Xinglong Station}

The optical spectroscopic observations were made on September 4-5,
2005, and January 03-08, 2006 at Xinglong station, Beijing
Astronomical Observatory (BAO). Low resolution spectra were taken
with the OMR spectrograph mounted on the BAO 2.16 m telescope, using
a Tektronix $1024 \times 1024 $ CCD as a detector. A
300~line~mm$^{-1}$ grating was used, blazed at 5200 \AA~for the
2005 September 04-05 and at 6000 \AA~for the 2006 January 03-08 runs,
respectively. Two exposures of 60 minutes each were taken for each
of the six nights at 2005 September 04-05, 2006 January03, 05, 07, and 08, and
one exposure of 60 minutes each for each of the two nights at
2006 January 04 and 06. The seeing ranged from $\sim 2^{''}.0-4^{''}.5$
during the observations, and we used slit widths of $2^{''}.0$ for
all of the observations when the seeing was less than $3^{''}.0$,
except for the three nights at 2006 January 03-05 when the seeing was the
largest ($\sim 4^{''}.5$). The resolution of the BAO spectra is
$\sim 10$ \AA~FWHM as measured from the night-sky lines. The CCD
reductions, including bias subtraction, flat-field correction, and
cosmic-ray removal, were accomplished with standard procedures using
IRAF. Wavelength calibration was carried out using Fe-He-Ar lamps
taken at the beginning and the end of the observations. The accuracy
of wavelength calibration was better than 1 \AA. Flux calibration
was derived with observation of a KPNO standard star, which was
observed at the beginning and end of each run.

Spectra obtained in the four nights with the best weather condition
(seeing $\sim 2^{''}.0$; 2005 September 04-05 and 2006 January 07-08)
 agree with
each other within $10\%$, indicating that 1) the uncertainty of our
flux calibration is better than 10\%, and 2) the source is not
variable on short (intranight) and medium (a few months)
time-scales. All of the spectra were corrected for the Galactic
extinction and combined to increase S/N ratio. The result is
displayed in Fig \ref{optspec}.

\subsection{SDSS and Other Spectra}

\begin{figure}
\epsscale{0.8} \plotone{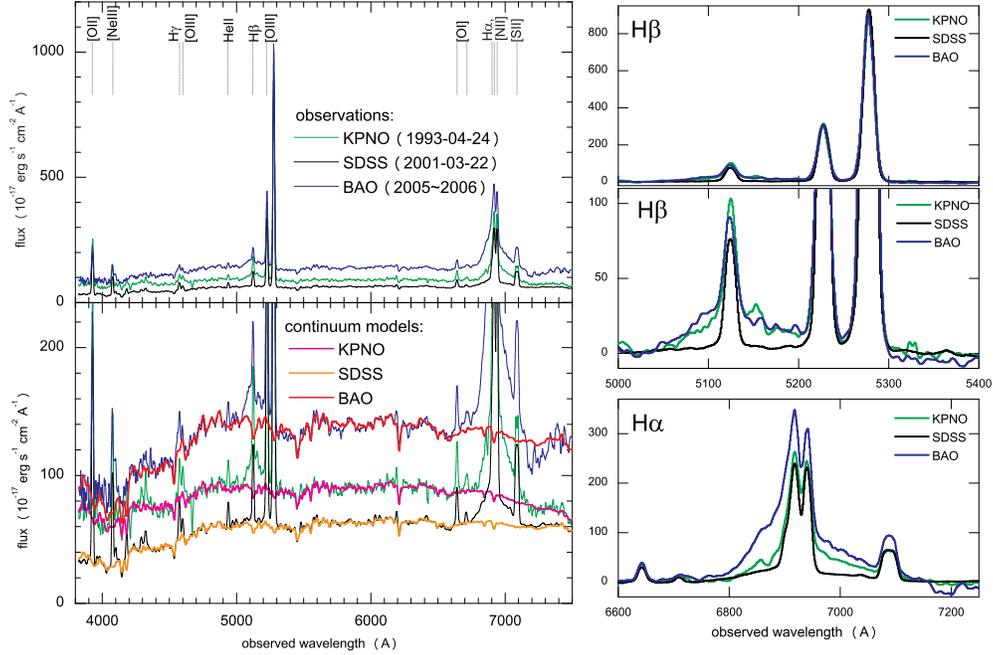} \caption{The optical spectra of Mrk
1393 taken at three epochs are normalized to [OIII]$\lambda$5007 and
shown in the upper-left panel (see text for details). The KPNO
spectrum is offset by $20\times 10^{-17}~erg~s^{-1}~cm^{-2}~\AA^{-1}$
for clarity. The lower-left panel is the same plot expanded in the
vertical axis for a clear view of the continuum and broad lines. Our
best fit continuum model is also over-plotted in the panel. Emission
line spectrum (after subtraction of the best fit continuum model) is
shown in the right panels (upper and middle panels for the
H$\beta$+[OIII] regime, and lower panel for the
[OI]+H$\alpha$+[NII]+[SII] regime). It is clear that the continuum
is dominated by starlight in all of the KPNO, SDSS, and BAO spectra.
Also note the dramatic variability of the broad H$\alpha$ and
H$\beta$ lines, while all of the narrow lines remain constant in the
three spectra. } \label{optspec}
\end{figure}

The SDSS spectrum of MRK 1393 was taken on 2001 March 22  with an
exposure time of 3041 s and a spectral resolution of about
$\lambda/\Delta\lambda =1800$ \citep{adelman07}. The seeing during
the exposure was around 2\farcs2. We started with the 1-d spectrum,
processed by the SDSS pipeline, and corrected it for the Galactic
reddening. The spectrum is also shown in Fig \ref{optspec}. Broad
H$\alpha$ appears much weaker than in the BAO spectrum, and the
broad H$\beta$, which can be easily identified in BAO spectrum, can
be hardly spotted in the original SDSS spectrum. The spectrum show
prominent stellar absorption lines, such as CaII, HCN, MgIb, NaID,
suggesting that the continuum is dominated by starlight.

A KPNO spectrum taken on 1993 April 24 was kindly provided by
Owen \citep{owen95}. The spectral resolution is 5.6 \AA~(FWHM). Owen
et al. estimated that the effective aperture size of the 1-d
spectrum is around $3^{''}$ in diameter with an uncertainty of less
than 30\%. The aperture includes about $\sim$95\% of the AGN light
(assuming a point-like source). The contribution of starlight should
be comparable to that in the SDSS spectrum. While both spectral
resolution and the signal-to-noise ratio of the KPNO spectrum are
lower than the SDSS spectrum, major stellar absorption lines, such
as CaII, MgIb, NaI can be easily identified (Fig \ref{optspec}).

Two more optical spectra can be found in the literature
\citep[][hereafter, MW88]{kinman83,mw88}. The spectrum taken in 1984
by MW88 shows prominent broad Balmer lines up to H$\epsilon$ and a
moderately blue continuum, which is very different from the SDSS
(2001) and the BAO (2005-2006) spectra. There is no observation time
in the Kinman's paper, though it must be taken before August 1981.
These authors also listed the emission lines they measured.

\subsection{Variations in the Continuum and Emission Lines}

In order to compare the strengths of emission lines among different
observations, we re-calibrated all the spectra using the 
[OIII]$\lambda$5007\AA\ line
flux, assuming it is not variable and 
the [OIII] emitting region is not spatially resolved. Constancy 
of [OIII] line flux over a time scale of ten years probably is a 
good approximation, 
however, the point-like source approximation for NLR is a concern. We 
noticed that the size (containing ~98\% line flux) of [OIII] emission line 
region for Mrk 1393 is 1 kpc or 0."9, according to the scaling relation 
of Bennert et al. (2002). This size is smaller than the spatial resolution, 
but is still 
comparable to the seeing disk and the aperture size used for the 
spectrum extraction, thus we warn that the absolute flux of 
the broad emission line and continuum flux may have uncertainties of order 
of ten percent. The [OIII]$\lambda$5007 emission line flux measured from the 
stacked BAO spectrum is only $83.5\%$ of
that measured from the SDSS spectrum. The difference can be ascribed
most likely to the aperture effect. The average seeing disk during
our observations was larger than that of the SDSS observation, while
the apertures we adopted to extract 1-d BAO spectra are less than
that for the SDSS spectrum (about $2^{''}.0\times 2^{''}.5$ for BAO
vs. $3^{''}$ in diameter for SDSS). Assuming [OIII] is not variable
over timescales of a few years, we multiply the BAO spectrum by a
factor of 1.198 to match the [OIII] flux measured in the SDSS to 
spectrum. We note that the [OI], [SII] and [NII] fluxes are
consistent with such a correction factor. The [OIII] flux measured
in the KPNO spectrum is a factor of 3.3 higher than that in the SDSS
spectrum, while the correction appears somewhat smaller in the blue
part as suggested by [OII]$\lambda$3727 flux (3.0). This can be explained 
by either effects of differential atmospheric refraction or more extended 
[OII] emission region. We divide the spectrum by 3.3.

After recalibrating the fluxes, we convolved the SDSS and KPNO
spectra with Gaussians to match the instrumental resolution of the
BAO spectrum. These spectra are displayed in Fig. \ref{optspec}. It
is evident that the strength of both the H$\alpha$ and H$\beta$
broad lines varied significantly between different epochs, and the
broad H$\beta$ line varied by an even larger factor, whereas all of
the narrow emission lines remain constant (note that we used only
[OIII]$\lambda$5007 as the calibrator). The continuum shapes and
intensities of the three spectra also appear to be quite different.
The continuum in the KPNO spectrum is only $\sim 12\%$ higher than
that in the SDSS spectrum, while the overall shape of the former is
significantly bluer than the latter; this is very likely due to
different nuclear contributions and/or different nuclear continuum
slopes.

The continuum modeling and emission line measurements were carried
out in the way as described in detail in \citet{z06} for the BAO,
SDSS and KPNO spectra. The host galaxy starlight is subtracted using
the templates constructed using the ICA method \citep{lu06}. The
nuclear component is described by a reddened power-law. The emission
lines are modeled with Gaussians: one Gaussian for each of the
narrow lines, and two Gaussians for each of the broad lines. We
assumed that the H$\alpha$ and H$\beta$ broad lines have the same
redshifts and profiles. Bad pixels are masked in modeling the
continuum and emission lines. This yields satisfactory fits to all
the data. The SDSS spectrum can be fairly well modeled by starlight
alone suggesting that the nuclear continuum is weak. In contrast, a
reddened power-law component is required in the BAO and KPNO
spectra.

\begin{figure}
\epsscale{0.8} \plotone{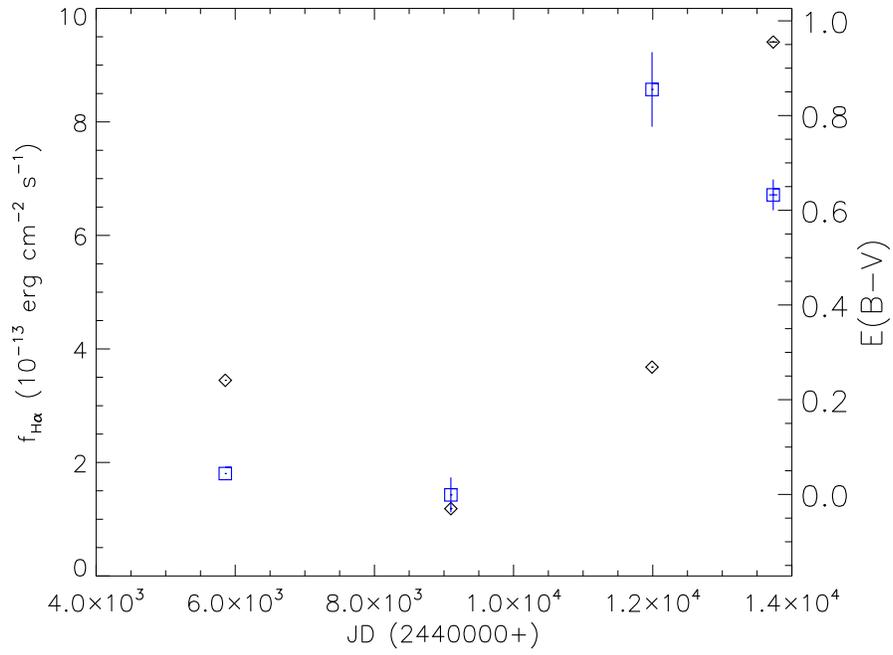} \caption{The overall variation of the
broad H$\alpha$ flux (``$\lozenge$'') and the reddening
(``$\square$'') estimated from the Balmer decrement. } \label{var}
\end{figure}

The Balmer decrement of the broad line component displayed large
variations on time-scales of several years. We found, for the broad
line component, H$\alpha$/H$\beta=5.95\pm$0.23 for the BAO spectrum
(taken at the end of 2005 and early 2006), 7.69$\pm$0.73 for the
SDSS spectrum (2001), and H$\alpha$/H$\beta=3.33\pm0.12$ for the
KPNO spectrum (1993). We also estimated the broad line Balmer
decrement to be 3.04 in 1984 using the integrated H$\alpha$ and
H$\beta$ line fluxes given by MW88, assuming a constant narrow
H$\alpha$, H$\beta$, [NII] line flux over ten years. This is
rectified because all other narrow line fluxes quoted in MW88 are in
good agreement with our measurement in the SDSS spectrum.  The flat
Balmer decrement is consistent with its blue continuum spectrum. The
overall variation of the broad H$\alpha$ flux and the intrinsic
reddening estimated from the Balmer decrement are plotted in Fig.
\ref{var}. The reddening is estimated from broad line Balmer
decrement, assuming an intrinsic value of H$\alpha$/H$\beta$=3.0
\citep{dong05,z06, dong08}.

\section{Discussion}

\subsection{A Reddening Origin for the Variable Balmer Decrement}

Dong et al. (2008) showed that the Balmer decrements for blue
quasars are in a very narrow range around 3.0 and do not correlate
with either the continuum slope, Eddington ratio, or luminosity.
Based on this and high luminosities in both the $K$-band and [OIII]
emission line, \citet{dong05} argued that objects with steep Balmer
decrements in their sample are intrinsically luminous objects 
reddened by dust. On the other hand, steepening of Balmer
decrement when continuum dimming has been reported for individual
objects \citep[e.g.,][]{wam90,tran92}. \citet{goo89} argued that the
variable Balmer decrement in some objects of their sample can be
attributed to dusty clouds moving into/out of the line of sight.

The variable Balmer decrement in Mrk 1393 can be most naturally
interpreted as variable extinction. First, X-ray absorption with a
column density around 2~4$\times 10^{21}$~cm$^{-2}$ by cold material,
as expected, has been detected in this source when it showed a steep
Balmer decrement of $H\alpha/H\beta \approx 6.0$. Second, the
observed broad H$\alpha$ flux during the BAO observation in
2005-2006 with a steep Balmer decrement is very similar to that in
1984 when the Balmer decrement was normal, indicating that the steep
Balmer decrement detected by BAO is not caused by continuum
luminosity variations as observed during the lower state of NGC
5548. Also consistent with continuum extinction is the fact that Mrk
1393 displayed a blue continuum in 1984 when the Balmer decrement is 
normal, while all of the reddened spectra show a stellar dominated 
continuum. We note that the stellar dominated continuum in 1993 is 
likely due to the intrinsic weakness of the continuum, since the 
H$\alpha$ broad line flux is about 4 times lower than in 1984.

Now that it has been established that the steep Balmer decrement in
Mrk 1393 is caused by dust reddening, we can use the Balmer
decrement to correct for the dust extinction. The extinction
corrected broad H$\alpha$ flux was the lowest in 1993 (at the KPNO
observation), and the highest in the end of 2005 and early 2006
(during our BAO observations), giving the largest observed variation
of a factor of 7.5. The intrinsic  H$\alpha$  flux was in between
these values at the epoch of the 1984 observation \citet{mw88} and
the 2001 SDSS observation (a factor of 1.9 and 2.1 higher than the
lowest state, respectively). These broad H$\alpha$ line flux
variations do not correlate with the Balmer decrements. X-ray flux 
during the ROSAT observation in 1991 is a factor of six lower than 
in the end 2005 and early 2006, perhaps indicating the object was in a low 
state in the early 1990s. We suggest that  Mrk 1393 is a good candidate 
for long-term monitoring. 
Because the Balmer decrement varies, as what was seen
from 1993 though 2001 to 2006, the object could be an ideal
laboratory for detailed studies of the physical and geometric
properties of obscuring matter.

\subsection{Dust-to-Gas Ratio of the obscuring gas}

Since the BAO spectra were taken only one and half month after the
\xmm\ observation and nearly {\em simultaneous} with the {\em Swift}
observation, it is reasonable to assume little or no variation for
the BLR reddening on such short time scales. The quasi-simultaneity
of the X-ray and optical observations enables us to derive the
dust-to-gas ratio for the obscuring gas in Mrk 1393. Assuming an
intrinsic H$\alpha$/H$\beta$=3.0, the average value for blue quasars
\citep{dong08}, Mrk 1393 was reddened by $E(B-V)$=0.59 during the
BAO observations. The column densities derived from the X-ray
absorption from the \xmm\ and {\em Swift} data are around
2.0--3.6$\times$10$^{21}$cm$^{-2}$ for different spectral models 
assuming solar abundances. We
find  a dust to gas ratio of $E(B-V)/N_H=1.6-3.0 \times 10^{-22}$
mag~cm$^{2}$ if the absorber is cold. In the two phase ionized 
absorber model for X-ray spectrum, the column density of the low 
ionization material was found to be 5.4$\times$10$^{21}$cm$^{-2}$, 
which gives a dust to gas ratio of 1.1$\times 10^{-22}$ mag~cm$^{-2}$.   
These values are within a factor of two of the average Galactic value 
1.7$\times 10^{-22}$ mag~cm$^{-2}$\citep{bohlin78}. Furthermore, 
the strong variability in reddening in Mrk 1393, as well as in several 
other intermediate Seyfert galaxies \citep{goo89}, warns against the use of
non-simultaneous optical and X-ray data to derive the dust-to-gas
ratio, at least for individuals.

As a comparison, the observed dust-to-gas ratios in other AGN show a
large diversity. A class of (low- and high-luminosity) AGN has
emerged which show relatively high optical-UV extinction, but lack a
corresponding amount of (cold) X-ray absorption (e.g., Brandt et al.
1997, Komossa \& Fink 1997b, Leighly et al. 1997, Komossa \& Bade
1998). Three mechanisms have been proposed to account for these
observations: (i) a higher-than-Galactic dust/gas ratio, (ii) a
different grain-size distribution with a dominance of small grains,
and (iii) dust mixed with absorbing material which is {\em ionized}
rather than neutral (Komossa \& Fink 1997b). In at least some 10 AGN
the third solution is favored (see Komossa 1999 for a review). On
the other hand, Maiolino et al. (2001) presented a sample of mostly
luminous Seyfert galaxies which show indications for dust/gas ratios
much lower than the Galactic value (by a factor 3-100). It is
possible that the observed diversity of the dust to gas ratios in
AGN can be reconciled by a combination of different obscuring
components.

\subsection{On the Nature of the Obscuring Gas}

The obscuring material seen in Mrk 1393 must be very close to the
AGN. By a comparison of the Balmer decrements of the broad and
narrow components, \citet{dong05} pointed out that the reddening of
the NLR is much smaller than that of the BLR in most of the
partially obscured quasars in their sample. The authors then
suggested that the obscuring material lies in between the BLR and
the NLR. The observed variations of the reddening set constraints to
the size to an even smaller region in Mrk 1393. The object was not
seriously reddened in 1993 (the KPNO observation) but was heavily
reddened in 2001 (the SDSS observation) and 2005-2006 (the BAO
observations). We interpret this as the obscuring matter moving into
($\sim $1993-2001) and out of ($\sim $2001-2006) the line-of-sight
to the nucleus (both of the continuum source and the BLR), as dust
condensation likely takes a much longer time in the vicinity of the
galactic nucleus. The size of the BLR, estimated from the H$\alpha$
line luminosity (reddening corrected) and the empirical $R_{BLR}$
versus H$\alpha$ luminosity relationship, is $R\sim$15 light days,
following \citet{wz03}. Using the revised relation given by Kaspi et
al. (2006) results $R\sim$19 light days. The obscuring material must
cover most of the BLR, and thus the transverse velocity of the
obscuring material must be not less than 1000 km~s$^{-1}$. With this
transverse velocity, the absorber is unlikely much farther than a
few parsecs away from the nucleus. As such,  it may be part of the
putative torus or even within it. If this is the case, our result
suggests that a dust-to-gas ratio similar to the Galactic value can
be found for gas located very close to the AGN. \citet{matt00} proposed
that Compton thin Seyfert 2/1.8/1.9 galaxies are absorbed by dust
lanes in their host galaxies. However, this is not the case for Mrk
1393.

Certain constraints on the gas properties can be inferred from the
data. Assuming that the reddening variability is caused by a single
cloud event and the cloud is quasi-spherical, we can set an
upper-limit on the particle density of the cloud of the order of
10$^6$ cm$^{-3}$, by using the X-ray absorption column density and a
cloud size no less than the BLR size. Alternatively, if the cloud is
pancake-like under the radiation pressure, the density of the gas
can be higher. The cloud is unlikely self-gravitational bound
because the tidal force of the central black hole exceeds the
self-gravity of the cloud by 1--2 orders of magnitudes for the 
above estimated density, if the distance to the black hole is less 
than 10 pc. The disrupted cloud will have a filament structure and the low 
column density gas will be blown away due the strong AGN radiation 
pressure. Thus the material must be either transient or bounded by 
some external pressure against tidal disruption.

Instead of appealing to the unknown external pressure, the transient
scenario is much more natural. One natural possibility is that the
absorbing gas is debris disrupted from the dusty torus. The black
hole mass of Mrk 1393 is estimated to be 5.5 $\times$10
$^7$\,M$_\odot$ from the M$_{bh}-\sigma$ relation \footnote{ We
measured a stellar velocity dispersion of 159$\pm$5 km~s$^{-1}$ in
the SDSS spectrum following the method of \citet{lu06}, which is
consistent with 163$\pm$15 km~s$^{-1}$ as given by Greene \& Ho
(2006). } \citep[e.g.][]{tre02}, and 2$\times 10^8$\,M$_\odot$ from
the above BLR size and the emission line width ($\sigma_{H\alpha}
=2940$~km~s$^{-1}$). The bolometric luminosity estimated
from the observed hard X-ray luminosity
(9$\times$10$^{43}$~ergs~s$^{-1}$) is about 0.10$-$0.37 of the
Eddington luminosity, assuming a bolometric correction to the 2-10
keV X-ray luminosity of 30 \citep{mar04}. Fabian et al.\ (2008)
showed that the radiation force exceeds the gravity of a black hole
for a dusty gas below the column density $N_H< 5\times
10^{23}\lambda$ cm$^{-2}$, when $\lambda$ is the Eddington ratio.
With $\lambda=0.10-0.37$, the measured X-ray column density in Mrk
1393 is much lower than this critical value, and thus the gas must
be accelerated outward by the radiation force if the black hole is
the only source of gravitational force. The masses of gas and stars
within a few parsecs do not likely exceed that of the black hole by
a large factor, so they are insufficient to prevent the outflow.
Thus, the obscuring gas we are seeing is either formed in the inner
nuclear region, or bound  and shielded by thick material
before it is exposed to the nuclear radiation. However, in the
presence of a strong radiation field, it is unrealistic to form dust
in relatively thin material within the inner parsecs. A plausible
origin of the gas is thus debris disrupted from the dusty torus.

To summarize, our XMM-Newton and Swift observations detected significant
X-ray absorption from cold gas with a column density of about
$2 - 4\times10^{21}$~cm$^{-2}$ in the partially obscured luminous
Seyfert galaxy Mrk 1393. Our semi-simultaneous BAO optical
spectroscopic observations gave a steep Balmer decrement of
$H\alpha/H\beta\approx 6.0$. A combination of the X-ray and optical
observations suggests that the steep Balmer decrement is caused by
dust reddening, and the dust-to-gas ratio of the obscuring material
is similar to the Galactic value. Incorporating the archived and
literature data, we find variations in Balmer decrement on a
time-scale of several years. Such variations require that the
obscuring material locates within a few parsecs from the central
nucleus. We suggest that the obscuring material
may be debris evaporated/disrupted from the dusty torus. The source 
also displayed large amplitude X-ray variations of a factor 6 
over 15 years.

\acknowledgements

We thank the anonymous referee for useful comments. We thank Stefanie 
Komossa for many helps in the XMM proposal and in writing the paper.
This work is supported by Chinese Natural Science Foundation through
CNSF-10573015, 10533050, an CAS knowledge innovation project
No. 1730812341 and national 973 project (2007CB815403). 
This work is partly based on observations obtained
with XMM-Newton, an ESA science mission with instruments and
contributions directly funded by ESA Member States and the USA (NASA).
Swift is supported at PSU by NASA contract NAS5-00136. This work has
made use of data obtained at Xinglong station, Chinese National
Astronomical Observatories, and data from SDSS.

\begin{figure}
\epsscale{0.8} \plotone{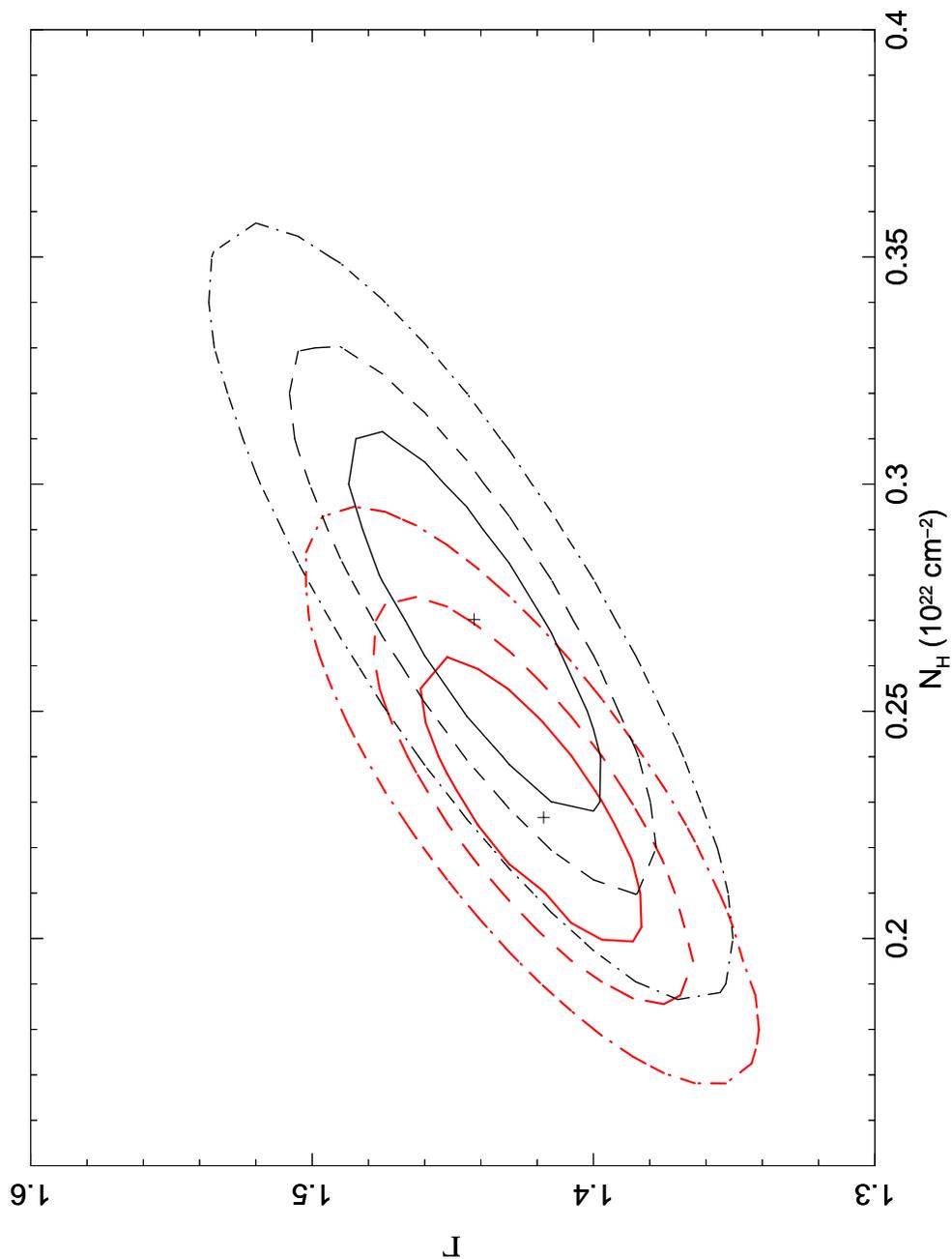} \caption{The contours at 70\%, 90\%
and 95\% ($\Delta\chi^2=$2.3, 4.6 and 9.2) confidence levels on the
plane of the intrinsic absorption column density $N_H$ versus photon
index for model-c (red) and model-d (black) in Fig 1. Model-e gave a
contour very similar to model-d, thus not show here for clarity. }
\label{contour}
\end{figure}

\begin{deluxetable}{clllllll}
\tablecaption{\label{xray_res} Results of acceptable X-ray spectral fits}
\tablehead{
\colhead{model} & \colhead{$N_H$} &
\colhead{$\Gamma$} & \colhead{$F_{2-10 keV}$} &
\colhead{k$T^{bb}/\Gamma_2$} & \colhead{$kT^{meka}$} &
\colhead{$F^{meka}_{0.5-2.0keV}$} & \colhead{$\chi^2$/dof} \\
\colhead{(1)} & \colhead{(2)} & \colhead{(3)} & \colhead{(4)} &
\colhead{(5)} & \colhead{(6)} & \colhead{(7)} & \colhead{(8)}
}
\startdata
\multicolumn{8}{c}{\xmm\ 1-10 keV} \\
1 & 2.42$_{-0.60}^{+0.56}$ & 1.50$_{-0.06}^{+0.05}$ & 8.17 & & & & 431/478 \\
\multicolumn{8}{c}{\xmm\ 0.3-10 keV} \\
2 &  3.02$_{-0.47}^{+0.46}$ & 1.53$_{0.05}^{+0.05}$  & 8.12 & & 0.166$_{-0.004}^{+0.004}$ & 5.69 & 667/650  \\
3 &  2.37$_{-0.03}^{+0.04}$ & 1.48$_{-0.04}^{+0.05}$ & 8.21  & 0.054$_{-0.011}^{+0.008}$ & 0.167$_{-0.005}^{+0.005}$ & 3.59 &  643/649  \\
4 &  2.25$_{-0.04}^{+0.03}$ & 1.48$_{-0.04}^{+0.04}$ & 8.10  & 5.69$_{-1.00}^{+0.97}$ & 0.168$_{-0.018}^{+0.006}$ & 3.09 & 634/650 \\
\multicolumn{8}{c}{\swift\ 1-10 keV}\\
1 & 3.3$_{-2.0}^{+1.1}$ & 1.58$_{-0.19}^{+0.21}$ & 7.5 & & & & 34/51 \\
\multicolumn{8}{c}{\swift\ 0.5-10 keV}\\
4 & 3.6$_{-2.2}^{+3.1}$ & 1.60$_{-0.21}^{+0.24}$ & 7.5 & 4.5$_{-2.9}^{+1.9}$ & 0.156$_{-0.078}^{+0.041}$ & 3.18 & 43/62
\enddata
\tablenotetext{a}{
model: Galactic absorption has been applied to all models. 
The absorption column density listed in this table is referred 
to the intrinsic  absorption at the redshift of Mkn 1393.
       1 -- a power-law absorbed by cold gas in the source rest frame. \\
       2 -- a power-law absorbed by cold gas plus thermal plasma
            emission ({\em meka} in {\em Xspec})
       3 -- a power-law and a black body absorbed by cold gas
            plus thermal plasma emission
            [$wabs\times (zwabs (powerlaw+zbb)+meka)$ in {\em XSPEC}].\\
       4 -- a broken power-law absorbed by cold gas
            plus thermal plasma emission [$wabs (zwabs\times bknpow+meka)$]
}
\tablenotetext{b}{(2) absorption column densities in units of
$10^{21}$~cm$^{-2}$; (4) modeled X-ray flux in units of $10^{-12}$~ergs~cm$^{-2}$~s$^{-1}$;
(5) k$T$ of black-body component in units of keV; (6) k$T$ of {\em meka} component in units
of keV;  (7) flux of {\em meka} component in 0.5-2.0 keV in units of $10^{-13}$~ergs~cm$^{-2}$~s$^{-1}$.  \label{table1} }
\end{deluxetable}

\end{document}